\documentclass[preprint,12pt]{elsarticle}
\usepackage{amsmath,amsthm,amsfonts,amssymb}

\def\eref#1{Eq.~(\ref{#1})}
\def\fref#1{Fig.~\ref{#1}}

\journal{International Journal of Non-Linear Mechanics}

\begin{document}

\begin{frontmatter}

\title{Non-smooth model and numerical analysis of a friction driven structure for piezoelectric motors}
\author[zju]{Weiting Liu}
\author[zju]{Maoying Zhou}
\author[zju]{Xiaodong Ruan}
\author[zju]{Xin Fu\corref{cor1}}
\ead{xfu@zju.edu.cn}
\cortext[cor1]{Corresponding author}
\address[zju]{The State Key Lab of Fluid Power Transmission and Control, Zhejiang University, Hangzhou, Zhejiang, China}

\begin{abstract}
  In this contribution, typical friction driven structures are summarized and presented considering the mechanical structures and operation principles of different types of piezoelectric motors. A two degree-of-freedom dynamic model with one unilateral frictional contact is built for one of the friction driven structures. Different contact regimes and the transitions between them are identified and analyzed. Numerical simulations are conducted to find out different operation modes of the system concerning the sequence of contact regimes in one steady state period. The influences of parameters on the operation modes and corresponding steady state characteristics are also explored. Some advice are then given in terms of the design of friction driven structures and piezoelectric motors.
\end{abstract}


\end{frontmatter}

\section{Introduction}
\label{introduction}
  Piezoelectric motors have been extensively studied and successfully applied to such areas as medical instruments \cite{zhao2011ultrasonic}, and consumer electronics \cite{ko2009piezoelectric} during the past few decades, with their outstanding features \cite{uchino1997piezoelectric} of high torque at low speed, quick response, quiet operation and compact size. They usually achieve mechanical output through the frictional contact between the between the stator and rotor/slider under the help of the induced elastic vibration in the composite stator. According to their operation principles, piezoelectric motors are divided into two types. In the first type of piezoelectric motors, resonant vibration is induced in the stator. Traveling wave piezoelectric motors \cite{sashida1985motor} and standing wave piezoelectric motors \cite{morita2000cylindrical} fall into this type. On the other hand, quasi-static vibration is usually induced in the stator of the second type of piezoelectric motors. As a result the rotor/slider usually conducts a quasi-static motion sequence somewhat like a stepping motor. Stick slip motors \cite{morita1999smooth} and inchworm motors \cite{kim2002hybrid} belong to this type.

  According to the underlying physics, mathematical modeling of piezoelectric motors mainly focuses on two parts: mathematical description of the electro-mechanical coupled stator and describing model for the contact interface between the stator and rotor/slider. The multi-physics nature of the coupled stator brings about complexity in terms of the coupling between elasticity and electricity. Some fruitful attempts have been made to manage the complexity using direct analytical method \cite{bouchilloux2003combined}, equivalent circuit method \cite{frayssignes2003traveling}, and finite element analysis method \cite{frangi2005finite}. The frictional contact between the stator and rotor/slider introduces strong nonlinearity and discontinuity into the model, especially when both the normal and the tangential contact are taken into consideration. Hitherto there have been numerous researches on the description of friction forces and their applications into piezoelectric motors. Hunstig et al \cite{hunstig2013stick1,hunstig2013stick2}  proposed a systematic analytical model for stick slip motors by adopting the Coulomb friction model, and investigated the influences of excitation signals on motor performances. Zharri \cite{zharii1993modeling} developed a mathematical model of a wedge-type ultrasonic motor considering the regime of slip and gave detailed discussions. Lu et al \cite{lu2000mathematical}  extended the model, divided each cycle of the stator vibration into several stages, and established the equations of rotor motion for each stage. Though the models provide insights into the operation of piezoelectric motors, little has been addressed in terms of the influences of contact properties on motor operation.

  In this contribution, two typical friction driven structures are put forward considering vibration of the stator and motion of the rotor in different piezoelectric motors. The second type of friction driven structure is modeled as a two degree-of-freedom mechanical system with one unilateral frictional contact. Dynamic equations of the system are developed in terms of the four identified frictional contact regimes. Simulations are conducted to recognize different operation modes of the friction driven structure regarding the number and order of different contact regimes in one steady state operation period. Furthermore, the influences of system parameters such as drive frequency, coefficient of friction and initial tilt angle of the converted rod on the steady state characteristics of the system are also evaluated.

\section{Friction driven structure and its dynamic model}
\label{dynamicmodel}
\begin{figure}
    \centering
    \includegraphics[width=\textwidth]{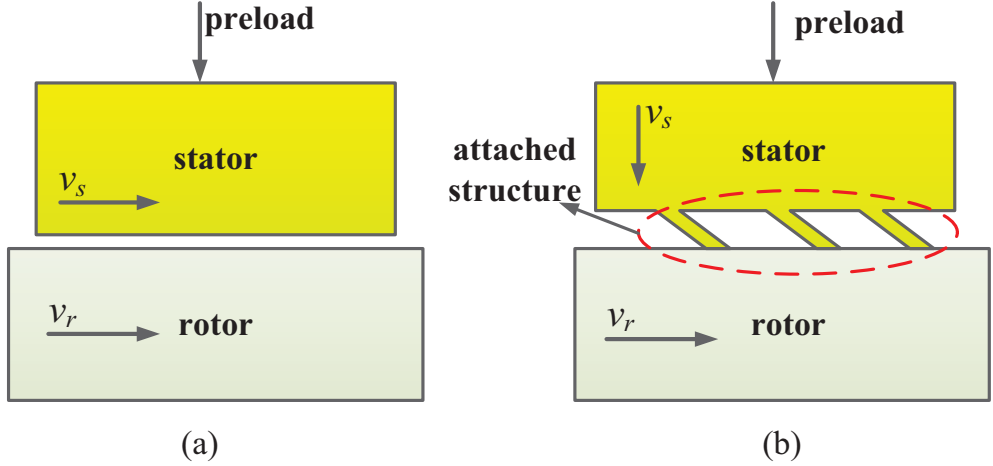}
    \caption{Friction driven structures: the first type (a), and the second type (b)}
    \label{fig_friction_drive}
  \end{figure}

  As described above, friction forces between the stator and rotor/slider play a key role in the operation of piezoelectric motors and largely determine the output performances, with the help of corresponding friction driven structures. According to different mechanical structures and operation principles of piezoelectric motors, two different types of friction driven structure can be extracted. For the first type, the direction of stator vibration is parallel to the resultant motion direction of rotor/slider, as shown in \fref{fig_friction_drive}(a). Traveling wave piezoelectric motors and stick slip motors belong to this type. Notice that in traveling wave motors, though the vibration direction of the mass particles in the stator is perpendicular to the rotor motion direction, the traveling wave excited in the stator propagates in a direction parallel to the rotor motion direction. For the second type including most standing wave motors and inchworm motors, direction of the stator vibration is usually different from that of the rotor motion, as shown in \fref{fig_friction_drive}(b). The generation of friction forces is achieved by the attached structure to the stator structure which couple the vibration of the stator and results in a tangential motion component.

\begin{figure}
    \centering
    \includegraphics[width=\textwidth]{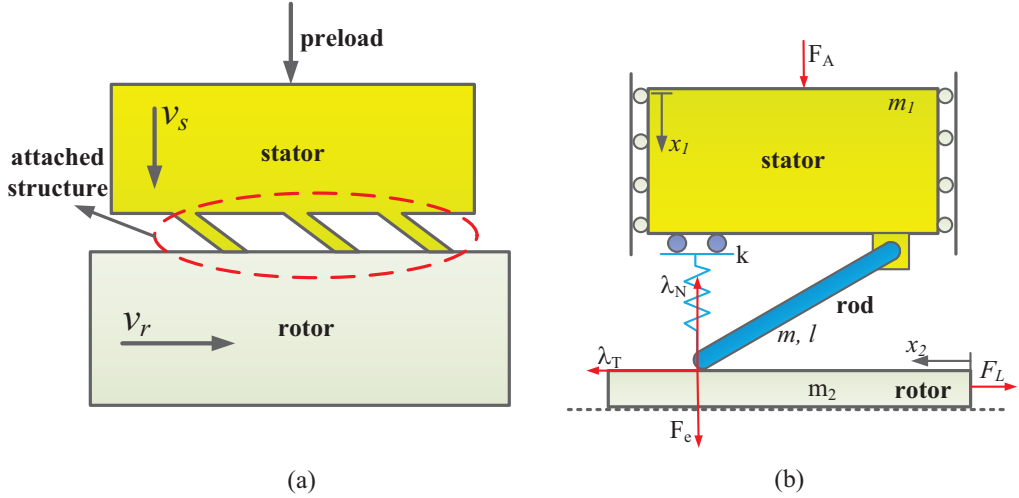}
    \caption{The second type of friction driven structure (a) and its simplified system (b)}
    \label{fig_friction_mechanism}
  \end{figure}

  The first type of friction driven structure is somehow similar to the belt conveyor system, and extensively studied. It is relatively simple as the normal forces in the frictional contact is constant. However, in the second type of friction driven structure, the normal contact forces are time varying, which brings about great complexity in the analysis and simulation. Hence in this paper we focus on the modeling and analysis of the second type of friction driven structure. As shown in \fref{fig_friction_mechanism}(a), the attached structure is usually in the form of cantilever beam and therefore can be replaced by a rigid link hinged to the stator with an extra supporting spring representing equivalent stiffness of the cantilever. As a result, the friction driven structure is converted the system shown in \fref{fig_friction_mechanism}(b), consisting of the stator, the rigid rod, the rotor block and the supporting spring.

  Choosing the vertical displacement $x_1$ of the stator block, the rotation angle $\varphi$ of the rod and the horizontal displacement $x_2$ of the rotor as the generalized coordinates of the system, the resultant kinetic and potential energy of the system are expressed as
  \begin{equation}
    \label{energy_term}
    \left\{ \begin{aligned}
      T &= \frac{1}{2}(m+m_1)\dot{x}_{1}^2 + \frac{1}{2}m_2\dot{x}_{2}^2 + \frac{1}{6}ml^2{\dot\varphi}^2 + \frac{1}{2}ml\dot\varphi\dot{x}_1\cos\varphi \\
      V &= \frac{1}{2}kl^2(\sin\varphi_0-\sin\varphi)^2
    \end{aligned}\right. ,
  \end{equation}
  where $m_1$ is the mass of the stator, $m$ is the mass of the rod, $m_2$ is the mass of the rotor, $l$ is the length of the rod, $k$ is the equivalent stiffness of the supporting spring, and $\varphi_0$ is the initial tilt angle of the rod. The normal and tangential contact forces between the rod and the rotor are denoted by $\lambda_N$ and $\lambda_T$ respectively and the external forces applied to the rotor $m_2$ and stator $m_1$ are represented by $F_L$ and $F_A$ respectively. The corresponding external work is
  \begin{equation}
    \label{work_done}
    \begin{aligned}
      W_e = & F_A(x_1-x_{10}) + F_L(x_2-x_{20}) + \\
            & \lambda_N(l\sin\varphi_0 - l\sin\varphi - x_1) + \lambda_T(x_2 - l\cos\varphi).
    \end{aligned}
  \end{equation}

  With the Lagrangian being $L=T-V+W_e$, dynamic equations of the simplified system turn out to be
  \begin{equation}
    \label{all_dynamic}
    \left\{ \begin{aligned}
       & (m+m_1)\ddot{x}_{1} + \frac{1}{2}ml\ddot\varphi\cos\varphi = F_A + \frac{1}{2}ml\dot{\varphi}^2\sin\varphi \\
       & m_2\ddot{x}_2=\lambda_T-F_L \\
       & \frac{1}{2}m\ddot{x}_1\cos\varphi + \frac{1}{3}ml\ddot\varphi = kl\cos\varphi(\sin\varphi_0-\sin\varphi)+\lambda_T\sin\varphi -\lambda_N\cos\varphi
    \end{aligned}\right. .
  \end{equation}

  According to the operation principle of piezoelectric motors, vertical displacement $x_1$ of the stator is proportional to the applied voltage $U(t)$, which is usually in the form
  \begin{equation}
    U(t)=\frac{U_{pp}}{2}(1-\cos wt),
  \end{equation}
  where $U_{pp}$ is amplitude of the applied voltage, $w$ is angular frequency of the applied voltage that can be calculated as $w=2 \pi f$ with f being frequency of the applied voltage. Thus vertical displacement $x_1$ of the stator kinematics are known and expressed by
  \begin{equation}
    \label{all_x1}
    \left\{ \begin{aligned}
       & x_1 = x_{10}- x_{10}\cos {wt} \\
       & \dot{x}_1 = wx_{10}\sin{wt} \\
       & \ddot{x}_1 = w^2x_{10}\cos{wt}
    \end{aligned}\right. ,
  \end{equation}
  where $x_{10}$ is the amplitude of the displacement. As a result, \eref{all_dynamic} becomes
  \begin{equation}
    \label{simple_dynamic}
    \left\{ \begin{aligned}
       & m_2\ddot{x}_2=\lambda_T-F_L \\
       & \frac{1}{2}m\ddot{x}_1\cos\varphi + \frac{1}{3}ml\ddot\varphi = kl\cos\varphi(\sin\varphi_0-\sin\varphi)+\lambda_T\sin\varphi -\lambda_N\cos\varphi
    \end{aligned}\right.
  \end{equation}

  Defining the dimensionless generalized coordinates
  \begin{equation}
    \label{dimensionless_coordinate}
    \left\{ \begin{aligned}
       & u_1 = \frac{x_1}{l} \\
       & u_2 = \frac{x_2}{l}
    \end{aligned}\right.
  \end{equation}
  and the dimensionless parameters
  \begin{equation}
    \label{dimensionless_parameter}
    \left\{ \begin{aligned}
       & \tau = wt \\
       & f_L = \frac{F_L}{m l w^2} \\
       & f_T = \frac{F_T}{m l w^2} \\
       & f_N = \frac{F_N}{m l w^2}
    \end{aligned}\right.
  \end{equation}
  above \eref{simple_dynamic} are reformulated as follows:
  \begin{equation}
    \label{simple_nondynamic}
    \left\{ \begin{aligned}
       & J\ddot{u}_2=f_T-f_L \\
       & \frac{1}{2}\ddot{u}_1\cos\varphi + \frac{1}{3}\ddot\varphi = P\cos\varphi(\sin\varphi_0-\sin\varphi)+ f_T\sin\varphi -f_N\cos\varphi
    \end{aligned}\right. ,
  \end{equation}
  where
  \begin{equation}
  \left\{ \begin{aligned}
       & J=\frac{m_2}{m} \\
       & P=\frac{k}{mw^2}
  \end{aligned}\right.
  \end{equation}
  are recollected constants. And with the introduction of two definitions that
  \begin{equation}
    \label{introduced_term}
    \left\{ \begin{aligned}
       & G_s = J\ddot{u}_2 + f_L \\
       & F_s = \frac{1}{2}\ddot{u}_1\cos\varphi + \frac{1}{3}\ddot\varphi - P\cos\varphi(\sin\varphi_0-\sin\varphi)
    \end{aligned}\right. ,
  \end{equation}
  \eref{simple_nondynamic} are reformulated as
  \begin{equation}
    \label{reduced_form}
    \left\{ \begin{aligned}
       & f_T = G_s \\
       & f_T\sin\varphi -f_N\cos\varphi = F_s
    \end{aligned}\right. .
  \end{equation}

\section{Contact model between the rod and rotor}
\label{contactmodel}
  \subsection{Different contact regimes}
  To model the frictional contact between the rod and rotor, two quantities $g_N$ and $g_T$ are introduced to denote the normal and tangential contact distance between the rotor and rotor, respectively. Together with corresponding contact velocity and contact accelerations, the contact kinematics are given by
  \begin{equation}
    \label{normal_kinematics}
    \left\{ \begin{aligned}
       & g_N = l\sin\varphi_0-l\sin\varphi-x_1 \\
       & \dot{g}_N = -\dot{x}_1 - l \dot\varphi\cos\varphi \\
       & \ddot{g}_N = l\dot{\varphi}^2\sin\varphi-l\ddot\varphi\cos\varphi - \ddot{x}_1
    \end{aligned}\right. ,
  \end{equation}
  and
  \begin{equation}
    \label{tangent_kinematics}
    \left\{ \begin{aligned}
       & g_T = x_2-l\cos\varphi \\
       & \dot{g}_T = \dot{x}_2 + l \dot\varphi\sin\varphi \\
       & \ddot{g}_T = l\dot{\varphi}^2\cos\varphi+l\ddot\varphi\sin\varphi + \ddot{x}_2
    \end{aligned}\right. ,
  \end{equation}
  respectively. It should be noted that under the assumption of impenetrability, negative contact distance is forbidden, indicating that $g_N \geq 0$. Again we refer to the previously adopted scheme for non-dimensionalization and suppose that
  \begin{equation}
  \left\{ \begin{aligned}
       & \delta_N = \frac{g_N}{l} \\
       & \delta_T = \frac{g_T}{l}
  \end{aligned}\right. ,
  \end{equation}
  Above \eref{normal_kinematics} and \eref{tangent_kinematics} are reformulated as follows:
  \begin{equation}
    \label{normal_nonkinematics}
    \left\{ \begin{aligned}
       & \delta_N = \sin\varphi_0-\sin\varphi-u_1 \\
       & \dot{\delta}_N = -\dot{u}_1 - \dot\varphi\cos\varphi \\
       & \ddot{\delta}_N = \dot{\varphi}^2\sin\varphi-\ddot\varphi\cos\varphi - \ddot{u}_1
    \end{aligned}\right. ,
  \end{equation}
  \begin{equation}
    \label{tangent_nonkinematics}
    \left\{ \begin{aligned}
       & \delta_T = u_2-\cos\varphi \\
       & \dot{\delta}_T = \dot{u}_2 + \dot\varphi\sin\varphi \\
       & \ddot{\delta}_T = \dot{\varphi}^2\cos\varphi + \ddot\varphi\sin\varphi + \ddot{u}_2
    \end{aligned}\right. .
  \end{equation}

  To specify contact force laws, the contact is discussed separately in the normal direction and tangential direction. In the normal direction, the contact state is divided into flight regime and continuous contact regime, while according to the contact state in the tangential direction the continuous contact regime is further divided into positive regime, negative regime and stick regime.

  The flight regime is characterized by a positive contact distance $\delta_N > 0$ and varnishing contact forces $f_N = 0$ and $f_T = 0$. Thus \eref{simple_nondynamic} reduce to two separated ordinary differential equations
  \begin{equation}
    \label{flight_dynamic}
    \left\{ \begin{aligned}
       & J\ddot{u}_2=-f_L \\
       & \frac{1}{2}\ddot{u}_1\cos\varphi + \frac{1}{3}\ddot\varphi = P\cos\varphi(\sin\varphi_0-\sin\varphi)
    \end{aligned}\right.
  \end{equation}
  with which the system status can be obtain through conventional integration process.

  The positive regime is characterized by varnishing normal contact distance, velocity and acceleration $\delta_N=0$, $\dot{\delta}_N=0$ and $\ddot{\delta}_N=0$,
  and positive tangential velocity $\dot{\delta}_T>0$, which means that
  \begin{equation}
    \label{expressions_varphi}
    \left\{ \begin{aligned}
       & \varphi = \arcsin (\sin\varphi_0 - u_1) \\
       & \dot\varphi = -\frac{\dot{u}_1}{\cos\varphi} \\
       & \ddot\varphi = {\dot\varphi}^2\tan\varphi-\frac{\ddot{u}_1}{\cos\varphi}
    \end{aligned}\right. ,
  \end{equation}
  and
  \begin{equation}
    \label{positive_force_relation}
    f_T = - \mu f_N.
  \end{equation}
  according to the Coulomb friction law.

  The contact forces are then given by
  \begin{equation}
    \label{positive_force}
    \left\{ \begin{aligned}
       & f_T = \frac{\mu F_s}{\cos\varphi+\mu\sin\varphi} \\
       & f_N = \frac{- F_s}{\cos\varphi+\mu\sin\varphi}
    \end{aligned}\right. .
  \end{equation}

  Given the initial states $u_{2p}$, $\dot{u}_{2p}$, $\varphi_p$, $\dot{\varphi}_p$ and $\tau_p$ of the positive regime, system states during positive regime can be calculated as follows:
  \begin{equation}
    \label{positive_state}
    \left\{ \begin{aligned}
       & \dot{u}_2 =  \dot{u}_{2p} + \int_{\tau_p}^{\tau} \frac{1}{J}(\frac{\mu F_s}{\cos\varphi+\mu\sin\varphi}-f_L)ds  \\
       & u_2 = u_{2p} + \int_{\tau_p}^{\tau} \dot{u}_2 ds
    \end{aligned}\right. .
  \end{equation}

  The negative regime is somehow similar to the positive regime with varnishing normal contact kinematics and negative tangential velocity $\dot{\delta}_T<0$, indicating that $f_T = \mu f_N$. Given  initial dynamic states $u_{2n}$, $\dot{u}_{2n}$, $\varphi_n$, $\dot{\varphi}_n$ and $\tau_n$ of the negative regime, the corresponding contact forces and system states are expressed as
  \begin{equation}
    \label{negative_force}
    \left\{ \begin{aligned}
       & f_T = \frac{- \mu F_s}{\cos\varphi-\mu\sin\varphi} \\
       & f_N = \frac{- F_s}{\cos\varphi-\mu\sin\varphi}
    \end{aligned}\right. ,
  \end{equation}
  and
  \begin{equation}
    \label{negative_state}
    \left\{ \begin{aligned}
       & \dot{u}_2 =  \dot{u}_{2n} + \int_{\tau_n}^{\tau} \frac{1}{J}(\frac{-\mu F_s}{\cos\varphi-\mu\sin\varphi}-f_L)ds  \\
       & u_2 = u_{2n} + \int_{\tau_n}^{\tau} \dot{u}_2 ds
    \end{aligned}\right. .
  \end{equation}

  At last in the stick regime, all the contact kinematic quantities varnish that
  \begin{equation}
    \label{stick_state}
    \left\{ \begin{aligned}
       & \delta_N = \dot{\delta}_N = \ddot{\delta}_N = 0  \\
       & \dot{\delta}_T = \ddot{\delta}_T = 0
    \end{aligned}\right. .
  \end{equation}
  while the contact forces are constrained by
  \begin{equation}
    \label{stick_relation}
    - \mu f_N \leq f_T \leq \mu f_N
  \end{equation}
  in which the  contact forces can be calculated according to \eref{reduced_form} :
  \begin{equation}
    \label{stick_force}
    \left\{ \begin{aligned}
       & f_T = G_s \\
       & f_N = G_s\tan\varphi-\frac{F_s}{\cos\varphi}
    \end{aligned}\right. ,
  \end{equation}
  and the corresponding contact kinematics are explicitly expressed in \eref{normal_nonkinematics} and \eref{tangent_nonkinematics}.

  \subsection{Transition between different contact regimes}
  With the four contact regimes identified and analyzed, it remains to determine the transitions between different regimes, that is to say, the critical states of the dynamic system where the contact regime transfers from one to another. According to the pre- and post- transition regimes, there are totally 12 different transition states.

  If the dynamic system is in flight regime before the transition, the critical state is marked by varnishing normal contact distance $\delta_N=0$. With further consideration, if the normal contact velocity is negative $\dot{\delta}_N < 0$, impact occurs and after impact only nonnegative normal contact velocity $\dot{\delta}_N \geq 0$ is admissible according to the impenetrability of rigid bodies. In the case of $\dot{\delta}_N > 0$, it is called the returning point where the system reaches the critical state but fails to transit to another contact regime. In the case of $\dot{\delta}_N = 0 $, the impenetrability indicates further nonnegative normal contact acceleration $\ddot{\delta}_N \geq 0$. Once again, $\ddot{\delta}_N > 0$ means another returning point while $\ddot{\delta}_N = 0$ indicates the transition to continuous contact regime, which should be further identified according to tangential kinematics.

  With positive tangential contact velocity $\dot{\delta}_T > 0 $, the transition goes to the positive regime, while a negative tangential velocity $\dot{\delta}_T < 0 $ indicates the transition to negative regime. For the case of varnishing tangential contact velocity $\dot{\delta}_T = 0 $, tangential contact acceleration $\ddot{\delta}_T $ should be further examined. $\ddot{\delta}_T > 0 $ leads to the transition to positive regime and $\ddot{\delta}_T < 0 $ guarantees a transition to negative regime, while $\ddot{\delta}_T = 0 $ means that the post-transition contact regime is stick.

  The most complicated part in the decision of critical states in the flight regime is the possible frictional impact. Some comprehensive work has been done on this subject \cite{stewart2000rigid} and several impact models \cite{pfeiffer2000multibody} have been put forward. However, since the physical mechanism of impact process is still unsatisfactorily understood, none of these impact models fulfill both physical accuracy and mathematical sophistication \cite{stronge1990rigid}.

  When the pre-transition contact regime of the system is positive regime, the transitions are divided into two types: the transitions to normal flight regime and the transitions to other tangential contact regimes including negative regime and stick regime. The critical state corresponding to the transition to flight regime is characterized by varnishing contact forces $f_N=0,f_T=0$, which means that $F_s=0$ according to \eref{positive_force}. In other words, the contact force $f_N$ keeps positive in the positive regime, reduces to zero at the transition, and remains varnished in the following flight regime. On the other hand, the transition to negative regime or stick regime is characterized by varnishing tangential contact velocity $\dot{\delta}_T=0$. It remains to determine the post-transition contact regime according to the tangential contact acceleration $\ddot{\delta}_T$. However, the tangential contact acceleration $\ddot{\delta}_T$ is coupled with the contact forces $f_N$ and $f_T$ nonlinearly by
  \begin{equation}
    \label{tangential_transition_condition}
    \left\{ \begin{aligned}
       & \ddot{\delta}_T > 0 \Leftrightarrow f_T = -\mu f_N \\
       & \ddot{\delta}_T < 0 \Leftrightarrow f_T = +\mu f_N \\
       & \ddot{\delta}_T = 0 \Leftrightarrow |f_T| \leq \mu f_N
    \end{aligned}\right.
  \end{equation}
  Hence it is not straightforward to evaluate the values of $\ddot{\delta}_T$, $f_T$ and $f_N$. According to  \eref{simple_nondynamic}, \eref{normal_nonkinematics}, and \eref{tangent_nonkinematics}, we can express the contact accelerations in terms of the contact forces as follows
  \begin{equation}
    \label{evaluate_forces}
    \left\{ \begin{aligned}
       & \ddot{\delta}_N = A f_N - B f_T + D \\
       & \ddot{\delta}_T = -B f_N +C f_T + E
    \end{aligned}\right. ,
  \end{equation}
  in which
  \begin{equation}
    \label{coefficient_expressions}
    \left\{ \begin{aligned}
       & A = 3 \cos^2\varphi \\
       & B = 3 \sin\varphi\cos\varphi \\
       & C = \frac{1}{J} + 3\sin^2\varphi \\
       & D = \frac{3}{2}\cos^2\varphi\ddot{u}_1 -3P\cos^2\varphi(\sin\varphi_0-\sin\varphi) + {\dot\varphi}^2\sin\varphi -\ddot{u}_1 \\
       & E = - \frac{3}{4}\sin {2\varphi}\ddot{u}_1 -\frac{3}{2}P\sin{2\varphi}(\sin\varphi_0-\sin\varphi) + {\dot\varphi}^2\cos\varphi -\frac{f_L}{J}
    \end{aligned}\right.
  \end{equation}
  are all known coefficients based on current dynamic states of the  system and satisfy
  \begin{equation}
    \label{coefficient_signs}
    A>0,\quad B>0,\quad C>0,\quad AC>B^2.
  \end{equation}

  Suppose that Coulomb friction model is adopted, we then calculate the contact forces according to the geometric method proposed by David Baraff \cite{baraff1993issues}. The main results are summarized as follows. When it holds that $D \geq 0$, then \eref{tangential_transition_condition} and \eref{evaluate_forces} lead to the solution
  \begin{equation}
    \label{trans_flight}
    \left\{ \begin{aligned}
       & f_N = f_T = 0   \\
       & \ddot{\delta}_N = D \\
       & \ddot{\delta}_T = E
    \end{aligned}\right.
  \end{equation}
  and the resultant contact regime after transition is flight regime. When $D < 0$, it remains to discuss the problem in several cases further. If
  \begin{equation}
    \frac{AE+BD}{B^2-AC} < \frac{\mu D}{A+\mu B}
  \end{equation}
  the solution turns out to be
  \begin{equation}
    \label{trans_positive}
    \left\{ \begin{aligned}
       & f_N = \frac{- D}{A+\mu B}   \\
       & f_T = \frac{\mu D}{A+\mu B} = - \mu f_N \\
       & \ddot{\delta}_N = 0 \\
       & \ddot{\delta}_T = \frac{AE+BD+\mu (BE+CD)}{A+\mu B} > 0
    \end{aligned}\right.
  \end{equation}
  and the critical state undergoes transition to the positive regime. If
  \begin{equation}
    \frac{\mu D}{A+\mu B} \leq \frac{AE+BD}{B^2-AC} \leq \frac{- \mu D}{A - \mu B}
  \end{equation}
  the solution is
  \begin{equation}
    \label{trans_stick}
    \left\{ \begin{aligned}
       & f_N = \frac{AC+BE}{B^2-AC}   \\
       & f_T = \frac{AE+BD}{B^2-AC}   \\
       & |f_T| \leq \mu f_N \\
       & \ddot{\delta}_N = \ddot{\delta}_T = 0
    \end{aligned}\right.
  \end{equation}
  and the transition goes to the stick regime. At last if
  \begin{equation}
    \frac{AE+BD}{B^2-AC} > \frac{- \mu D}{A - \mu B}
  \end{equation}
  the solution becomes
  \begin{equation}
    \label{trans_negative}
    \left\{ \begin{aligned}
       & f_N = \frac{-D}{A - \mu B}   \\
       & f_T = \frac{- \mu D}{A - \mu B} = \mu f_N  \\
       & \ddot{\delta}_N =0 \\
       & \ddot{\delta}_T = \frac{AE+BD-\mu(BE+CD)}{A+\mu B} < 0
    \end{aligned}\right.
  \end{equation}
  and the corresponding transition goes to negative regime.

  When the system is in negative regime before transition, the results are much like those of the positive regime. In fact, we can obtain the results by replacing $\mu$ with $-\mu$ in the case where the pre-transition contact regime is positive regime.

  In the last place, if the system stays in stick regime before transition, the transition to flight regime is also defined by varnishing contact forces $f_N=f_T=0$, while the transitions to positive regime and negative regime are determined by two indicators $f_T- \mu f_N $ and $f_T+ \mu f_N$, respectively. When $f_T- \mu f_N$ increases to zero and $f_T+ \mu f_N$ remains positive, the transition goes to negative regime. If $f_T- \mu f_N$ remains negative while $f_T+ \mu f_N$ decreases to zero and further becomes negative, the transition goes to positive regime.

\begin{figure}[htbp!]
    \centering
    \includegraphics[width=0.7\textwidth]{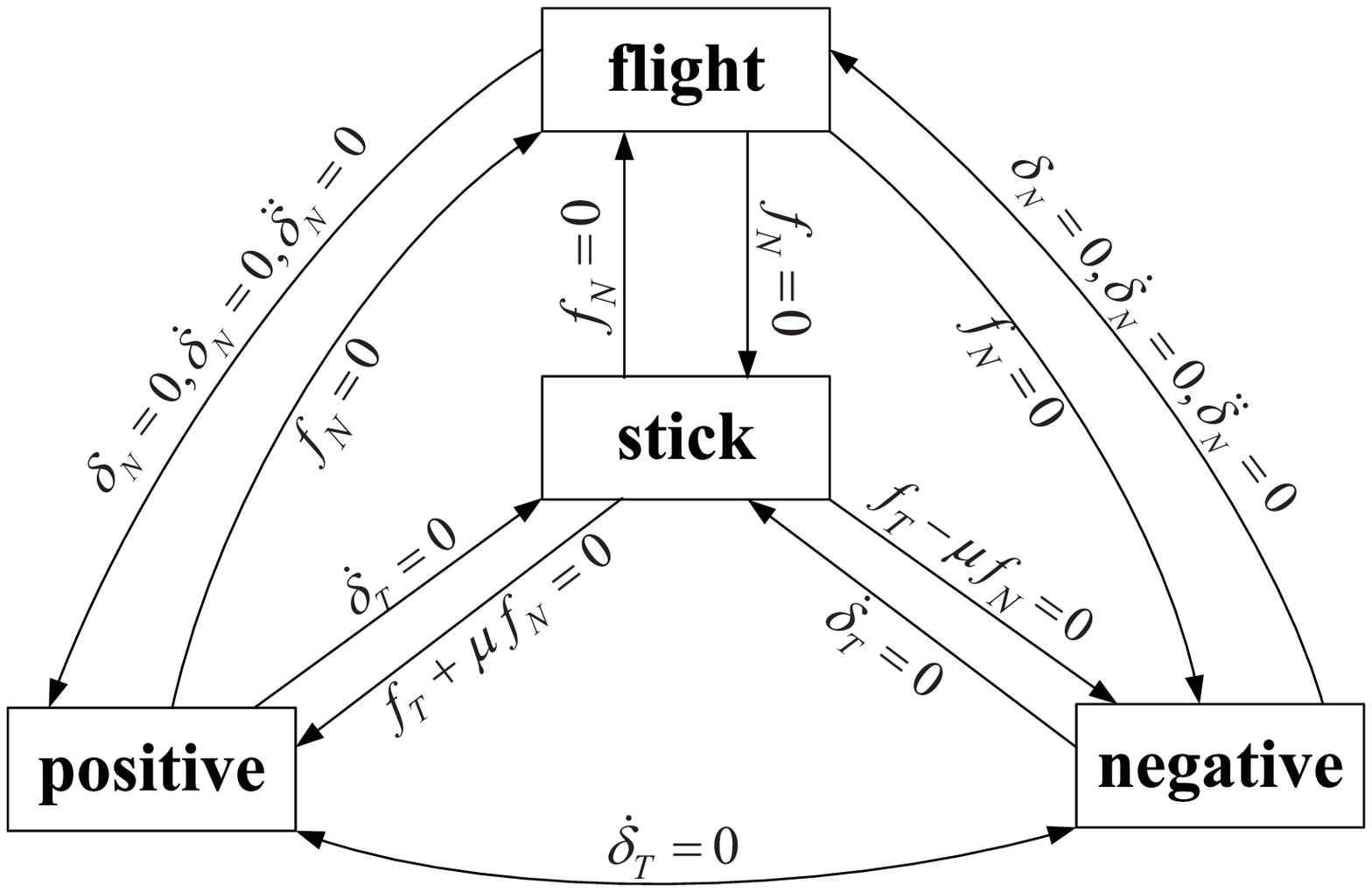}\\
    \caption{Transition between different regimes}\label{fig_transition_points}
  \end{figure}

  All in all, the above descriptions can be expressed in the  diagram shown in \fref{fig_transition_points}.

  \subsection{Painlev\'{e} paradox}
  Take the case of negative regime for an example, where $f_T= \mu f_N$. Combine this identity with \eref{evaluate_forces}, we obtain the following equation
  \begin{equation}
    \label{painleve_equation}
    \ddot{\delta}_N = (A - \mu B )f_N + D.
  \end{equation}

  When the rod is in contact with the moving rotor, the normal contact force $f_N$ should be greater than zero while normal contact acceleration $\ddot{\delta}_N$ must be zero, or $f_N$ can be zero while $\ddot{\delta}_N$ should be greater than zero. Thus we obtain the following linear complementarity problem (LCP)
  \begin{equation}
    \label{lcp_negative}
    \ddot{\delta}_N \geq 0, \quad f_N \geq 0, \quad \ddot{\delta}_N f_N = 0.
  \end{equation}

  The following cases are possible:
  \begin{enumerate}[1)]
     \item $A - \mu B  > 0$, there could be only one solution
       \begin{enumerate}[a)]
         \item $D>0$ implies that $\ddot{\delta}_N=0$, $f_N=D>0$ (flight regime);
         \item $D<0$ implies that $\ddot{\delta}_N=0$, $f_N=\frac{-D}{A- \mu B}$(positive regime or negative regime according to the sign of $f_T$);
       \end{enumerate}
     \item $A - \mu B  < 0$, there could be two solutions or no solutions
        \begin{enumerate}[a)]
           \item $D>0$ implies that $\ddot{\delta}_N=0$, $f_N=D > 0$ (flight regime), or $\ddot{\delta}_N=0$, $f_N=\frac{-D}{A-\mu B}$ (positive regime or negative regime according to $f_T$);
           \item $D<0$ implies no solutions exist.
        \end{enumerate}
  \end{enumerate}

  The so called Painlev\'{e} paradox refers to the case where the theoretical solution to the unilateral frictional contact problem loses existence and uniqueness, just as described above in the case of $A-\mu B <0$. Thus in the following analysis, we will always admit the following condition
  \begin{equation}
    \label{mu_constraint}
    A - \mu B > 0 \Leftrightarrow 1 - \mu\tan\varphi > 0.
  \end{equation}

\section{Numerical simulations}
\label{numericalsimulation}

  \subsection{Numerical simulation}
  According to the analysis in the last section, numerical simulation process of the proposed system can be summarized as follows:
  \begin{enumerate}[1)]
    \item Update the system states: $u_2$, $\dot{u}_2$, $\varphi$ and $\dot\varphi$; \label{step1}
    \item Evaluate the contact kinematics $\delta_N$, $\dot{\delta}_N$, and $\dot{\delta}_T$;
    \item If $\delta_N > 0$, the system is in flight regime, \eref{flight_dynamic} holds and system states can be updated using numerical integration method. After this, we go back to step \ref{step1});
    \item If $\delta_N = 0$ and $\dot{\delta}_N N 0$, impact occurs at the time instance. The impact laws proposed in references \cite{stewart2000rigid,pfeiffer2000multibody,stronge1990rigid} are adopted to determine the post-impact state $u_2$ , $\dot{u}_2$, $\varphi$ and $\dot\varphi$ of the system as well as the contact kinematics $\delta_N$, $\dot{\delta}_N$, and $\dot{\delta}_T$;
    \item If $\delta_N=0$ and $\dot{\delta}_N>0$, again the system proceeds to flight regime and \eref{flight_dynamic} is utilized to update system states. The simulation then goes to step \ref{step1});
    \item If $\delta_N=0$, $\dot{\delta}_N=0$ and $\dot{\delta}_T>0$, the system operates in positive regime and \eref{positive_state} is used to determine system states. The updated simulation jumps to step \ref{step1});
    \item If $\delta_N=0$, $\dot{\delta}_N=0$ and $\dot{\delta}_T<0$, the system is in negative regime and \eref{negative_state} determines the state of the system. The updated simulation jumps to step \ref{step1});
    \item If $\delta_N=0$, $\dot{\delta}_N=0$ and $\dot{\delta}_T=0$, the geometric method proposed by Baraff \cite{baraff1993issues} is used to solve \eref{evaluate_forces} and determine the contact forces $f_N$, $f_T$ as well as the contact accelerations $\ddot{\delta}_N$, $\ddot{\delta}_T$;
    \item With $\ddot{\delta}_N > 0$ and $f_N=f_T=0$, the system transits to flight regime and is governed by \eref{flight_dynamic}. A series of integration is done to update the system states and the simulation jumps to step \ref{step1});
    \item With $\ddot{\delta}_N = 0$, $\ddot{\delta}_T>0$ and $f_T=-\mu f_N$, the system transits to positive regime and the system states are updated according to \eref{positive_state}, after which the simulation jumps to step \ref{step1});
    \item With $\ddot{\delta}_N = 0$, $\ddot{\delta}_T<0$ and $f_T=\mu f_N$, the system transits to negative regime and the system states are updated according to \eref{negative_state}. As a result, the simulation goes back to step \ref{step1});
    \item With $\ddot{\delta}_N = 0$, $\ddot{\delta}_T=0$ and $-\mu f_N \leq f_T \leq \mu f_N$, the system transits to stick regime and the system states are updated according to \eref{stick_state}, after which the simulation jumps to step \ref{step1}).
  \end{enumerate}

  Typically, two types of integration scheme are used to deal with above simulation process, the event driven scheme and the time stepping scheme. \cite{brogliato2008numerical,leveque1992numerical} The event driven scheme \cite{lancioni2009non} is based on the decomposition of dynamic system operation into continuous operation modes and discrete transition events. Great effort of this scheme is made to detect and discriminate discrete transition events, meaning the 12 transition states in our problem. Between the discrete transition events, it is easy to calculate the system states according to the differential equations of the system in the corresponding operation modes, which means the four contact regimes in our problem. As a comparison, the time stepping scheme \cite{stewart1996implicit} is developed based on the theory of measure differential inclusions \cite{moreau1988unilateral} and complementarity problems \cite{lotstedt1982mechanical}. The whole numerical simulation process is conducted with fixed time step length. During each time step, the problem of the system states determination is generalized as a nonsmooth one-step problem, in which the system is modeled at velocity using differential inclusions and reformulated into a series of linear complementarity problems that can be solved by conventional algorithms.

  Though the time stepping scheme shows some advantages over the event driven scheme, such as smooth integration process and no events accumulation, it actually suffers from some instability caused by the solution of complex complementarity problem and fails to detect the accurate time for discrete events in an efficiently way. Usually, the time stepping scheme is applied to systems with multiple degrees of freedom and multiple unilateral contacts. Here in our case, the event driven scheme has proven to be a better choice.

  The parameters used in the numerical simulations are listed as follows:
  \begin{enumerate}[1)]
    \item dynamic parameters: $m=2.0 \times {10}^{-5} kg$, $m_2=2.0 \times {10}^{-3} kg$, $k=1.0 \times {10}^4 N/m$;
    \item geometry parameters: $l=1.0 \times{10}^{-3} m$, $\varphi_0=\pi/4$;
    \item contact parameters: $\mu=0.1$;
    \item force parameters: $x_{10}=2.0 \times{10}^{-6} m$, $F_L=0.0N$.
  \end{enumerate}

  In the simulation, we would like to explore the influences of system parameters $\varphi_0$, $\mu$ and drive frequency $f$ on the steady state operation of the proposed system. Besides, to avoid impact accumulations in the simulation process, we assume that the normal contact is always fulfilled with \eref{expressions_varphi}. This is achieved by constraining the values of parameters $\varphi_0$ and $\mu$ at given drive frequency $f$ with $f_N>0$ and $D<0$ fulfilled all the time. During the simulation, steady state operation of the simplified system is evaluated in terms of the constituent contact regimes, and the net displacement of the rotor in each operation period is calculated as well as average moving velocity of the rotor.

  \subsection{Different operation modes}
\begin{figure}[htbp!]
    \centering
    \includegraphics[width=\textwidth]{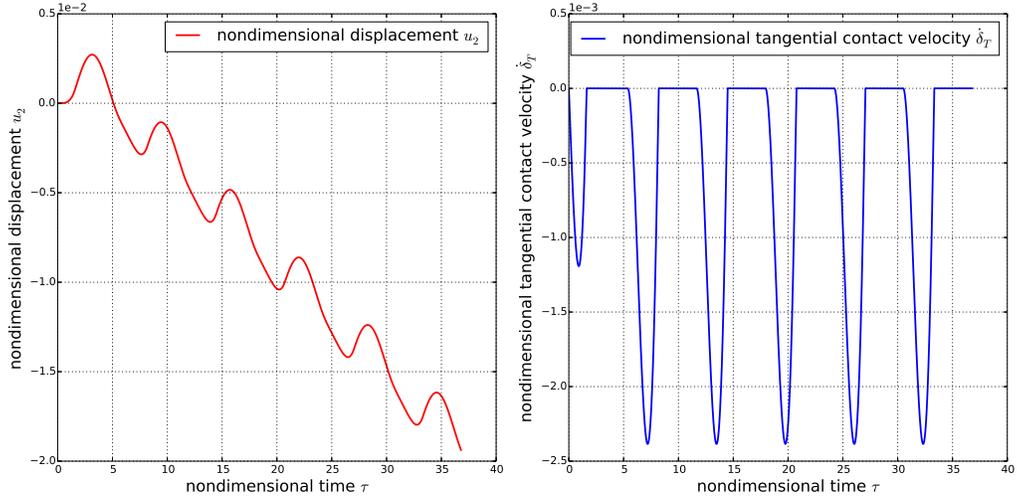}\\
    \caption{Operation mode NS: non-dimensional displacement (left) and tangential contact velocity (right) versus non-dimensional time}\label{fig_steady_state_01ns}
  \end{figure}

\begin{figure}[htbp!]
    \centering
    \includegraphics[width=\textwidth]{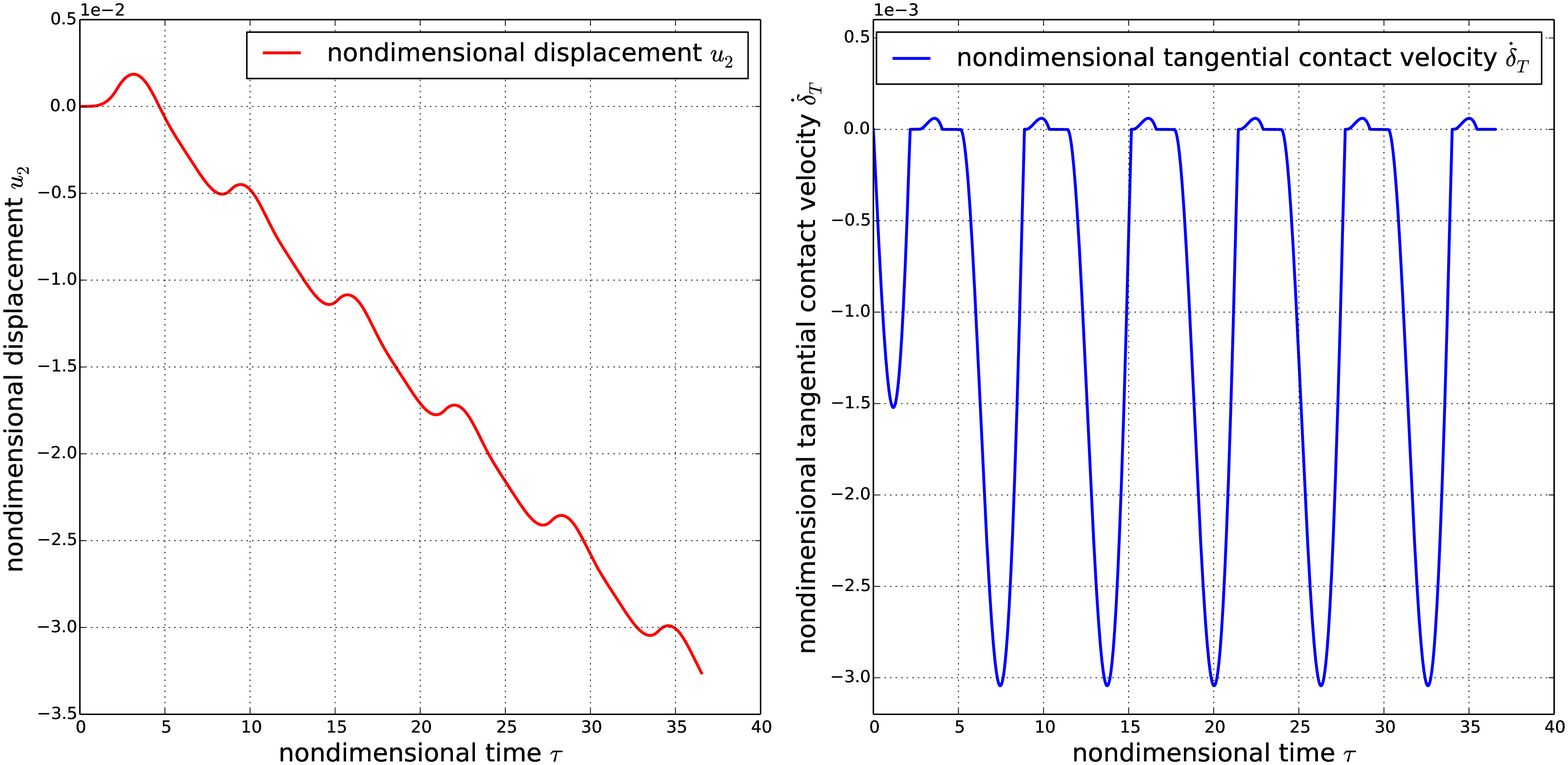}\\
    \caption{Operation mode NSPS: non-dimensional displacement (left) and tangential contact velocity (right) versus non-dimensional time}\label{fig_steady_state_02nsps}
  \end{figure}

\begin{figure}[htbp!]
    \centering
    \includegraphics[width=\textwidth]{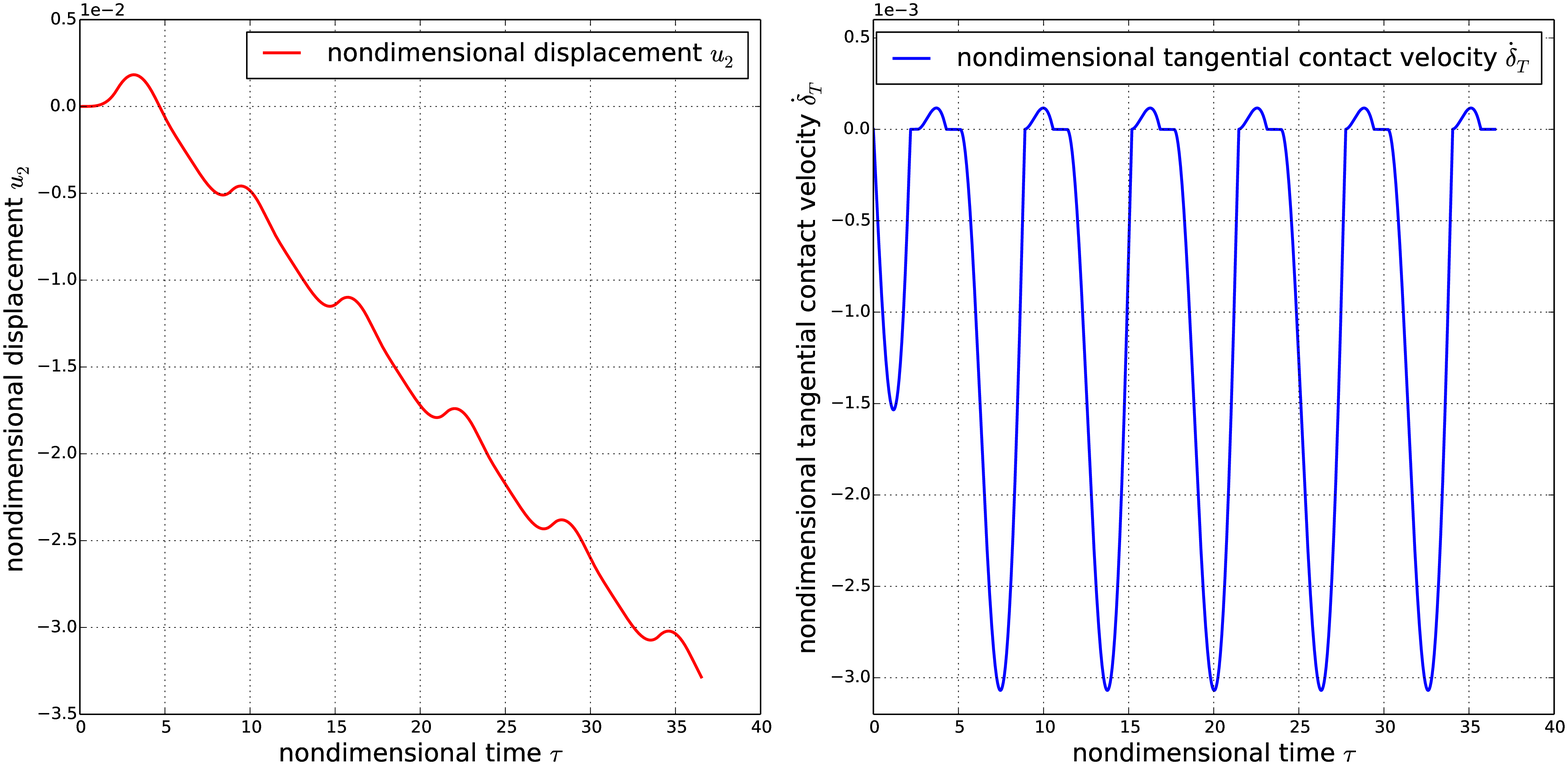}\\
    \caption{Operation mode NPS: non-dimensional displacement (left) and tangential contact velocity (right) versus non-dimensional time}\label{fig_steady_state_03nps}
  \end{figure}

\begin{figure}[htbp!]
    \centering
    \includegraphics[width=\textwidth]{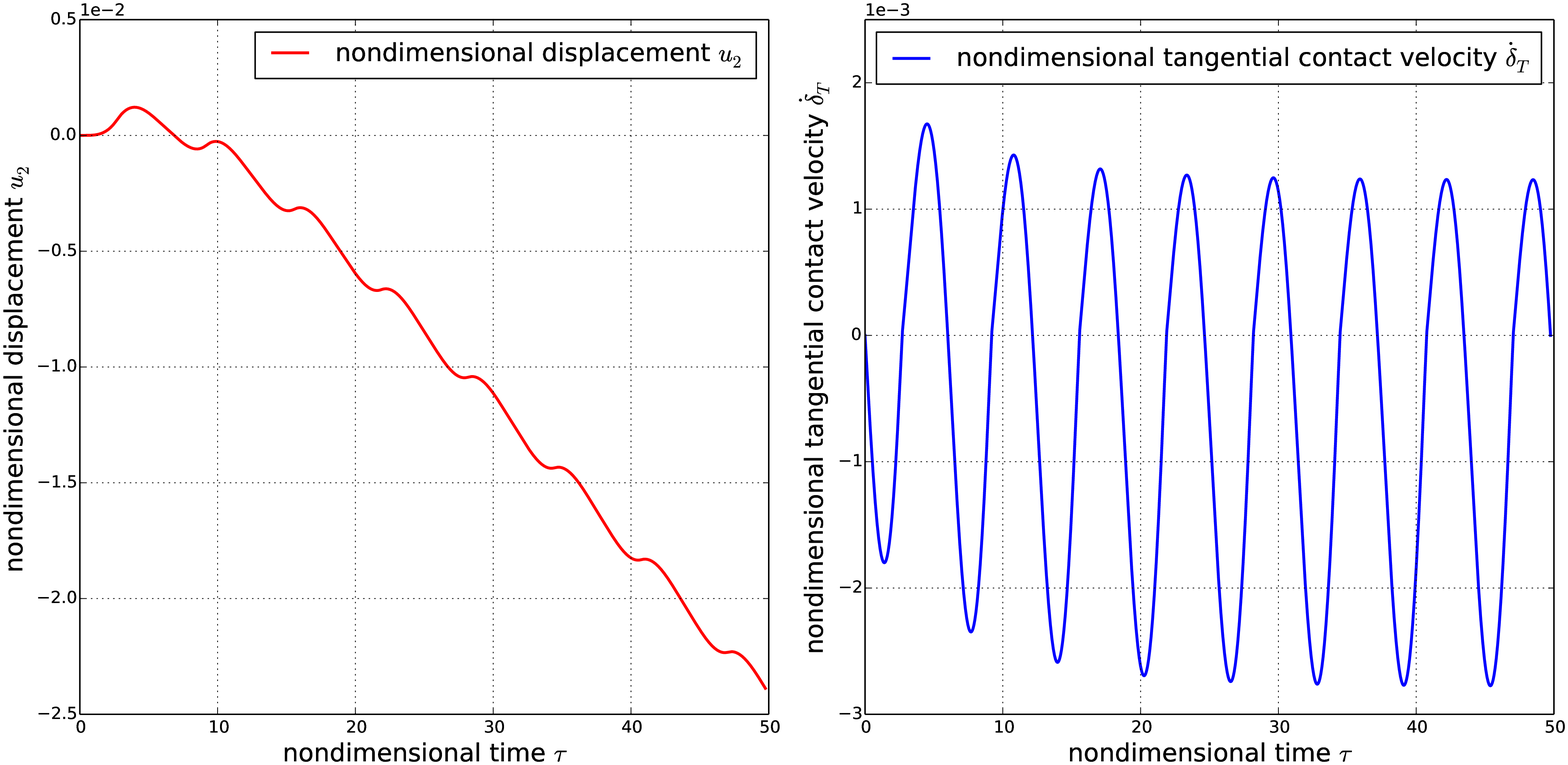}\\
    \caption{Operation mode NP: non-dimensional displacement (left) and tangential contact velocity (right) versus non-dimensional time}\label{fig_steady_state_04np}
  \end{figure}

\begin{figure}[htbp!]
    \centering
    \includegraphics[width=\textwidth]{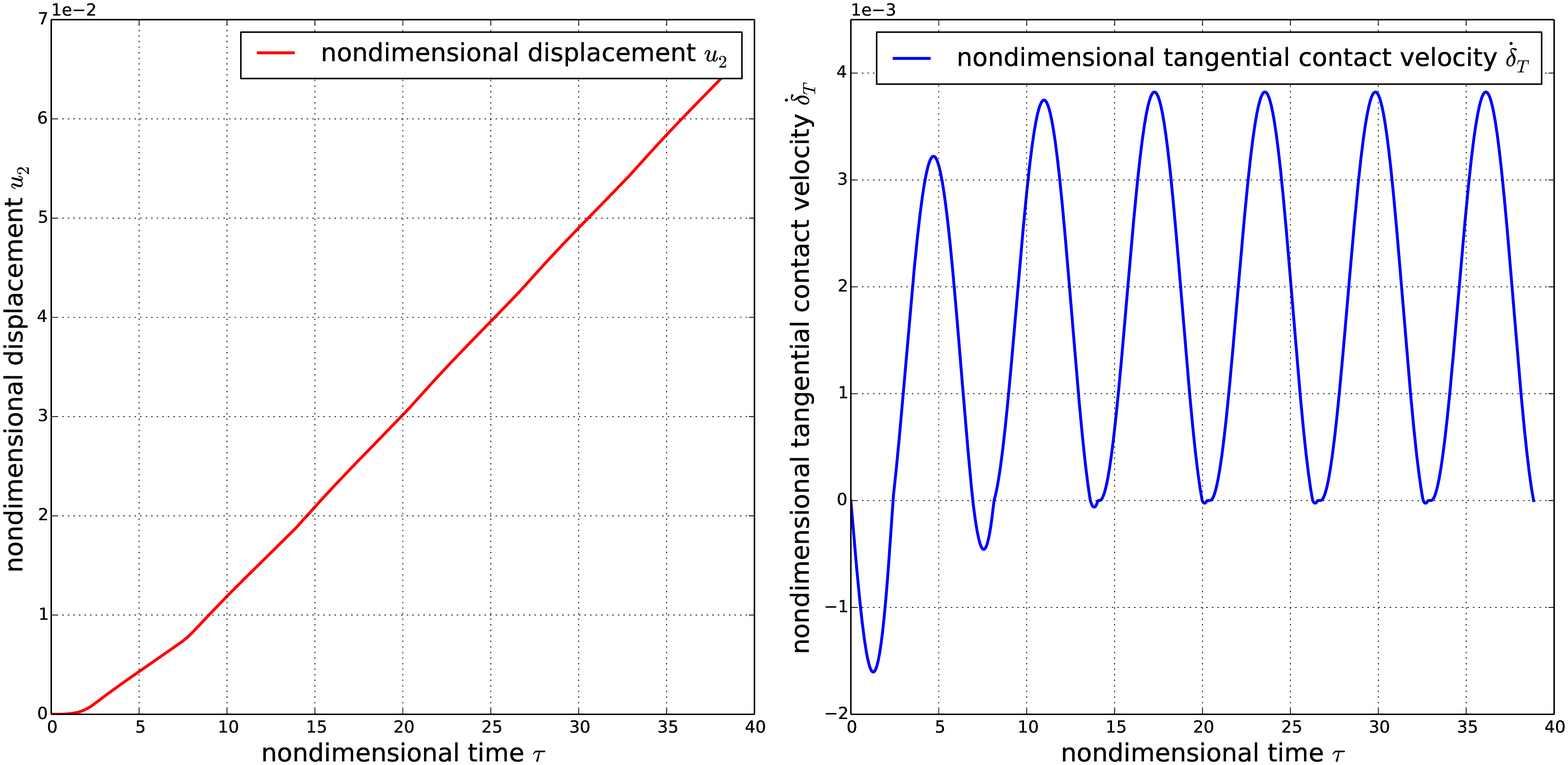}\\
    \caption{Operation mode NSP: non-dimensional displacement (left) and tangential contact velocity (right) versus non-dimensional time}\label{fig_steady_state_05nsp}
  \end{figure}

\begin{figure}[htbp!]
    \centering
    \includegraphics[width=\textwidth]{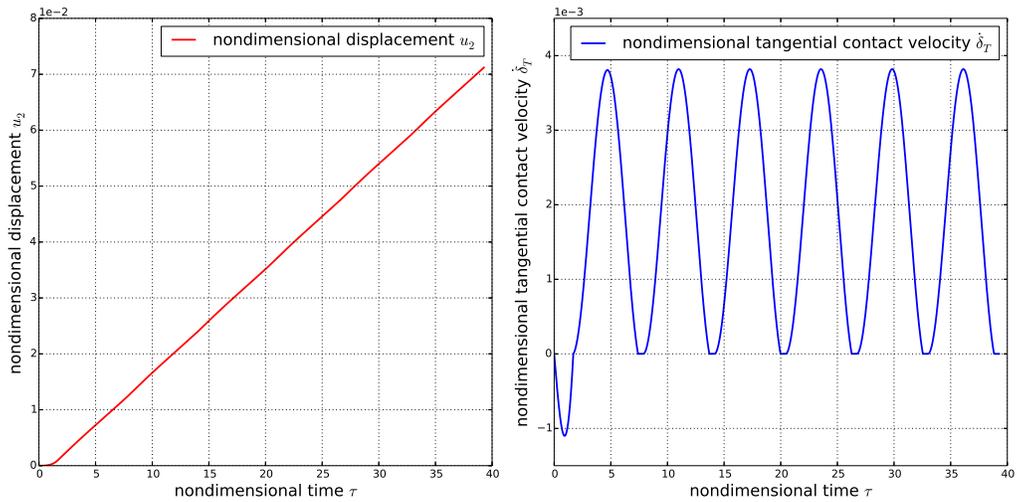}\\
    \caption{Operation mode SP: non-dimensional displacement (left) and tangential contact velocity (right) versus non-dimensional time}\label{fig_steady_state_06sp}
  \end{figure}

  According to the simulation results, six operation modes of the system are identified according to the sequence of the contact regimes contained in each steady state operation period:
  \begin{enumerate}[1)]
    \item Negative regime 〞 stick regime, as shown in \fref{fig_steady_state_01ns}. It is denoted by NS;
    \item Negative regime 〞 stick regime 每 positive regime 每 stick regime, as shown in \fref{fig_steady_state_02nsps}. It is denoted by NSPS;
    \item Negative regime 每 positive regime 每 stick regime, as shown in \fref{fig_steady_state_03nps}. It is denoted by NPS;
    \item Negative regime 每 positive regime, as shown in \fref{fig_steady_state_04np}. It is denoted by NP;
    \item Negative regime 每 stick regime 每 positive regime, as shown \fref{fig_steady_state_05nsp}. It is denoted by NSP;
    \item Stick regime 每 positive regime, as shown in \fref{fig_steady_state_06sp}. It is denoted by SP.
  \end{enumerate}

  According to the displacement curves shown above, for each steady state operation mode, the rotor can be seen as conducting stepping motion. The motion accuracy is relative to the net displacement of rotor in each steady state operation period. At the same time, motion direction of the rotor may be positive or negative. In fact, when the operation mode is NS, net motion of the rotor is negative and the rotor achieves its maximum motion velocity in this operation mode. Nevertheless if the operation mode is SP, the rotor shows a positive net displacement, the maximum positive motion velocity of the rotor is achieved in this operation mode. In the case where the operation mode is NP, the rotor may undergo positive or negative motion according to the drive frequency. (Note that in \fref{fig_steady_state_04np} only the case of negative net motion is shown)

  \subsection{Motion characteristics}
\begin{figure}[htbp!]
    \centering
    \includegraphics[width=0.75\textwidth]{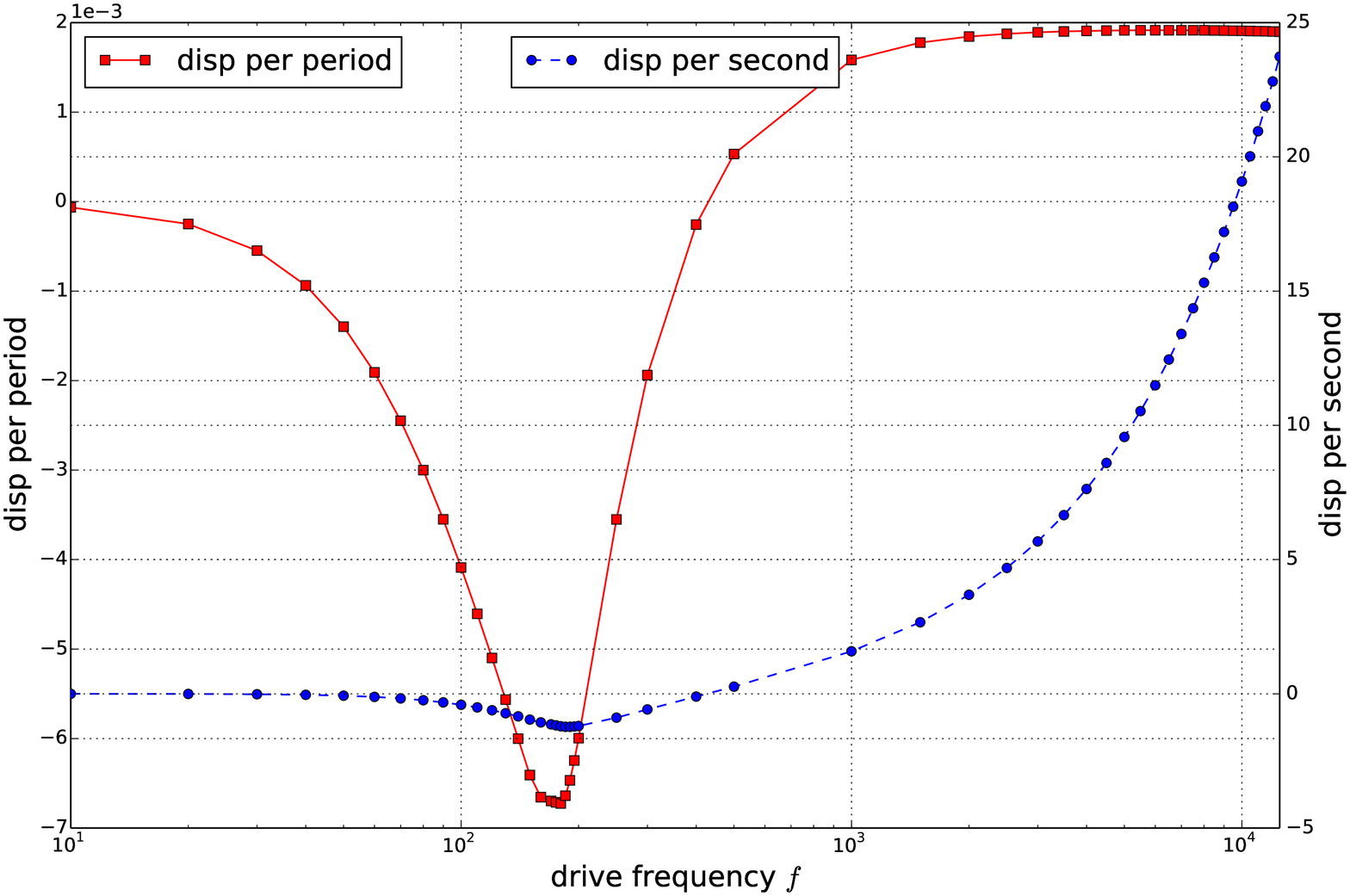}
    \caption{Motion characteristics of the rotor with varying drive frequency f}
    \label{fig_results_fr}
  \end{figure}

  Firstly, the drive frequency \emph{f} is varied from 10 Hz to 12500 Hz. With determined steady state operation mode at each applied frequency, net displacement of the rotor during one steady state operation period (disp per period) is evaluated and average velocity of the rotor (disp per second) is thus obtained. The results are plotted versus drive frequency and shown in \fref{fig_results_fr}. According to the results, with the increase of drive frequency, steady state operation mode of the simplified system varies from the above described first operation mode to the third operation mode and then to the fourth operation mode. The net displacement of the rotor per operation period reaches a minimum value at the drive frequency of 180 Hz, tends to zero with the drive frequency approaches zero and arrives at its maximum while the drive frequency approaches infinity. As a result, average velocity of the rotor is approximately proportional to the drive frequency when the frequency is relatively high, which validates the assumptions and modeling work presented previously \cite{zhou2014bio,liu2015modeling}.

  Secondly, coefficient of friction 米 varies from 0.01 to 0.99 and numerical simulations are conducted with drive frequency f being 20 Hz, 185 Hz and 2000 Hz, respectively. The net displacement of the rotor per operation period is evaluated and plotted versus coefficient of friction 米, as shown in \fref{fig_results_fr}. When the drive frequency is 20 Hz, steady state operation mode of the simplified system is always the first operation mode. And as 米 increases, the net displacement of the rotor per operation period remains negative and eventually approaches zero. When the drive frequency is 185 Hz, steady state operation mode of the simplified system varies from the fourth operation mode, to the third operation mode, to the second operation mode and finally to the first operation mode. The net displacement of the rotor per operation period remains negative and reaches its minimum with a coefficient of friction being 0.13. When the drive frequency is 2000 Hz, steady state operation mode of the simplified system varies from the fourth operation mode, to the fifth operation mode and finally to the sixth operation mode. The net displacement of the rotor per operation period keeps positive and approaches its maximum with coefficient of friction 米 being 0.89.
  \begin{figure}[htbp!]
    \centering
    \includegraphics[width=0.75\textwidth]{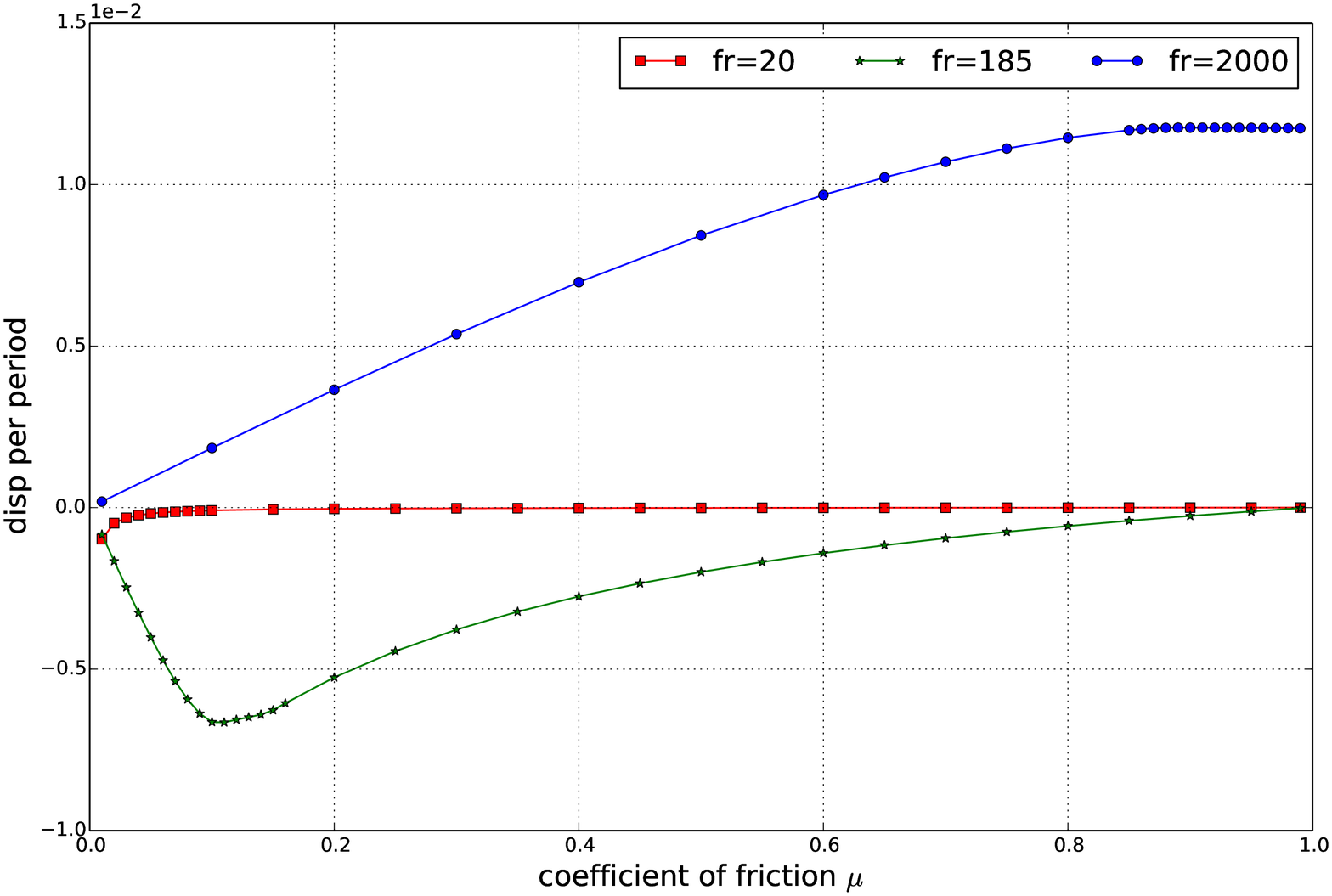}
    \caption{Motion characteristics of the rotor with varying coefficient of friction $\mu$}
    \label{fig_results_mu}
  \end{figure}

\begin{figure}[htbp!]
    \centering
    \includegraphics[width=0.75\textwidth]{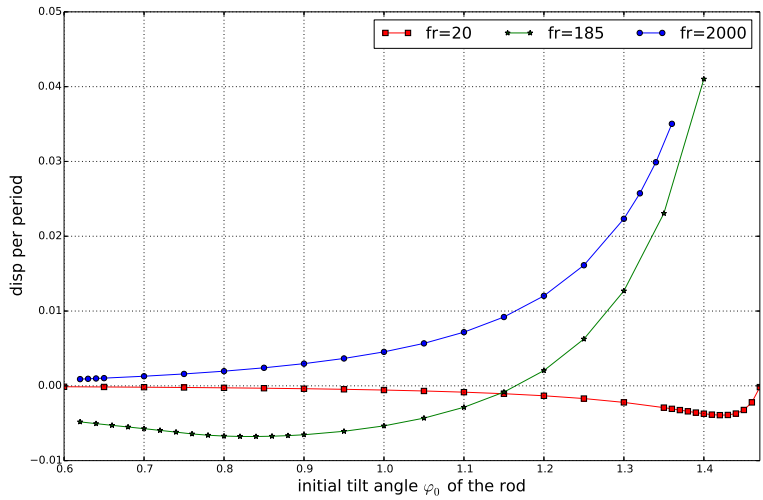}
    \caption{Motion characteristics of the rotor with varying initial tilt angle $\varphi_0$ of the rod}
    \label{fig_results_ph}
  \end{figure}

  Thirdly, numerical simulations are conducted with drive frequency f being 20 Hz, 185 Hz and 2000 Hz respectively and varying initial tilt angle $\varphi_0$ of the rod. The net displacement of the rotor per operation period is evaluated and plotted versus initial tilt angle $\varphi_0$ of the rod, as shown in \fref{fig_results_ph}. When the drive frequency is 20 Hz, steady state operation mode of the simplified system is always the first operation mode. The corresponding net displacement of the rotor per operation period keeps negative and reaches its minimum when $\varphi_0$ is 1.42. When the drive frequency is 185 Hz, steady state operation mode of the simplified system varies from the third operation mode to the fourth operation mode. The corresponding net displacement of the rotor per operation period reaches its minimum with $\varphi_0$ being 0.84. When the drive frequency is 2000 Hz, steady state operation mode of the simplified system is always the fourth mode and the net displacement of the rotor per operation period increases monotonously with $\varphi_0$.

\section{Conclusions and discussions}
\label{conclusion_discussion}
  In this contribution, two typical friction driven structures are put forward considering structures and operation principles of different piezoelectric motors. A two degree-of-freedom dynamic model is set up for the second friction driven structure. Different contact regimes and the transitions between them are identified and analyzed. Numerical simulations are conducted to find different operation modes of the system according to the sequence of contact regimes in one steady state period. The influences of system parameters on the steady state operation characteristics of the system are also explored. Nevertheless, several points should be noted.

  Firstly, in the design of friction driven structure, or further any kind of piezoelectric motors utilizing the friction driven structure, motion direction of the rotor is of great importance and largely determined by the steady state operation mode. For a positive velocity output, the SP operation mode is the best choice and for a negative velocity outpu, the NS operation mode is preferable. When bidirectional motion is required, the NP operation mode should be chosen. With steady state operation mode of the system determined, values of the system parameters are actually constrained and can be further optimized.

  Secondly, in the simulation we take the assumption that no impact events occur during system operation, which is not the case in realistic operation. Nevertheless, in the design of motor structure, impact events can be totally eliminated through the choices of appropriate parameters, as done in the simulation process and shown in \eref{mu_constraint}. Besides, the impact events bring about energy loss due to inelastic contact and thus should be avoided to improve power efficiency of the motor.

  Finally, the model presented above for the friction driven structure is still a preliminary one. Further improvements can be made by considering the higher vibration mode of the tilt rod and the nonlinearity introduced by piezoelectric elements, such as hysteresis and saturation. Besides, the system can be extended by adding another tilt rod with opposite tilting direction to the previous one and incorporating the two groups of the beam branches described in previous contribution\cite{zhou2014bio}. As a result, the extended system can be used to describe the behaviors of a number of other proposed piezoelectric actuators \cite{friend2006simple,friend2003single,uchiki1991ultrasonic,racine1998flexural,kurosawa1989ultrasonic,fleischer1989ultrasonic}.

\section*{Acknowledgements}
  This work has been developed under the support of the National Basic Research Program of China (No. 2011CB013303), International Science and Technology Cooperation Program of China (2012DFG71860) and the Science Fund for Creative Research Groups of National Natural Science Foundation of China (No. 51221004).

\section*{References}
\bibliographystyle{elsarticle-num}
\bibliography{maureenchou}
\end{document}